\documentclass[12pt]{article}
\usepackage[T2A]{fontenc}
\usepackage{amsmath, amsfonts, amssymb}
\usepackage{graphicx}
\usepackage{hyperref}
\usepackage{color}

\renewcommand{\Im}{\mathop{\mathrm{Im}}\nolimits}
\renewcommand{\Re}{\mathop{\mathrm{Re}}\nolimits}
\newcommand{\calC}{\mathcal C}
\newcommand{\calE}{\mathcal E}
\newcommand{\calF}{\mathcal F}
\newcommand{\bs}{\boldsymbol}

\author{W. Wachowski$^{a}$\thanks{vladvakh@gmail.com} \and P.\,I. Pronin$^{b}$\thanks{petr@phys.msu.ru}}
\title{Heat kernel for higher-order differential operators in Euclidean space}
\date{\it $^a$ I.\,E. Tamm Department of Theoretical Physics, P.\,N. Lebedev Physical Institute, Leninsky ave. 53, 119991 Moscow, Russia. \\
$^b$ Department of Theoretical Physics, Faculty of Physics, M.\,V. Lomonosov Moscow State University, 119991 Moscow, Russia.}

\begin {document}
% ======================================

\maketitle

\abstract{We consider heat kernel for higher-order operators with constant coefficients in $d$-dimensio\-nal Euclidean space and its asymptotic behavior. For arbitrary operators which are invariant with respect to $O(d)$-rotations we obtain exact analytical expressions for the heat kernel and Green functions in the form of infinite series in Fox--Wright psi functions and Fox $H$-functions. We investigate integro-differential relations and the asymptotic behavior of the functions $ \calE_{\nu, \alpha}(z)$, in terms of which the heat kernel of $O(d)$-invariant operators are expressed. It is shown that the obtained expressions are well defined for non-integer values of space dimension $d$, as well as for operators of non-integer order. Possible applications of the obtained results  in quantum field theory and the connection with fractional calculus are discussed.}

\tableofcontents

%%%%%%%%%%%%%%%%%%%%%%%%%%%%%%%%%%
\section{Introduction}
%%%%%%%%%%%%%%%%%%%%%%%%%%%%%%%%%%

The investigation of classical equations of mathematical physics and their applications is based on the consideration of the behavior of their fundamental solutions, the Green functions of the corresponding linear differential operators. In quantum field theory (QFT), the Green functions (the elementary particle propagators) also play a fundamental role. Due to perturbation theory and renormalization the Green functions allowed to calculate vacuum expectation values  of  fundamental fields products (which are also commonly called Green functions in QFT and correlation functions in statistical physics), the effective action of the theory and the quantities observed in the experiment, for example, the scattering cross sections.

However, in the middle of the last century, due to the work of Minakshisandaram \cite{Minakshi, Minakshi2}, Schwinger \cite{Schwinger} and DeWitt \cite{Dewitt}, it is became clear that it is convenient to calculate both the various physical quantities and the Green functions themselves using a new object --- ``heat kernel'', depending on the additional parameter --- ``proper time'' (see sect. \ref{S1}). The success of this new computational algorithm \cite{Seeley, Gilkey, Gilkey2} was due to the fact that, for the Laplace operator in flat background, the heat kernel really is the fundamental solution of the heat equation, so the well-known function (\ref{HeatKernel}). This allows us to construct the fundamental invariants of differential operators --- the HaMiDeW-coefficients (Hadamard--Minakshisundaram--DeWitt, or heat kernel coefficients).

Now the heat kernel method is one of the most powerful tools in mathematical physics, that has applications in a wide region from pure mathematics (spectral geometry) to analysis of financial markets. Being combined with the background field method in QFT it allows calculating the effective action and investigating the renormalizability of theories, the presence of anomalies in them, etc directly in the coordinate space. This makes it indispensable for computations in the presence of external fields or in curved space-time, which is crucially important for gauge theories and quantization of gravity \cite{Gibbons, JackOsborn, JackParker, LeeRim, Barvinsky85, Barvinsky87, CPT2, CPT4, Barvinsky03}. See also \cite{Avram00, Avram01, Vassil03} and references there.

 The heat kernel method also can be used in the investigation of higher order operators. This is important for regularization by means higher covariant derivatives, as well as for theories with higher derivatives that have attracted much interest in recent years, namely, $R^2$-gravity \cite{AvramBarvinsky}, nonlocal and superrenormalizable theories \cite{Tomb, Modesto} and Ho\u{r}ava-Lifshitz type theories \cite{Barvinsky17, Mamiya14}. One of the possible extensions of the standard heat kernel method was proposed in \cite{Barvinsky85}. It consists in deformation of higher order operators  to minimal operators (i.e. powers of an operator of the Laplace type), which makes it possible to use the usual second order heat kernels and HaMiDeW-coefficients. Discussion on the application of the heat kernel method for higher-order operators can also be found in \cite{Avram97, Avram98}.

Nevertheless, the heat kernels of higher-order differential operators are important quantities themselves. So the above mentioned possible application of the standard method to higher-order operators does not make it less interesting to study their heat kernels directly. Before the consideration of operators on manifolds, we have to investigate the behavior of heat kernel in Euclidean space. This article is aimed on the solution of this problem.

Section \ref{S1} is introductory. We describe heat kernel $U_F(\tau; \bs{x})$ of a differential operator $F(\nabla)$ in Euclidean space, propose the integral representations for the Green functions and consider the application of the standard method to calculation of the Green function $G_{\Delta^\nu}(\bs{x})$. In section \ref{S2}, the heat kernel of the operator $-(-\Delta)^\nu$ is obtained in two different ways. The explicit form of $U_{\nu, d}(\tau; \bs{x})$ and newly introduced functions $\calE_{\nu, \alpha}(z)$ are considered in section \ref {S3}. We represent them in terms of the Fox--Wright psi functions and discuss their domain, their connection with the theory of differential equations of fractional order and the limit for $\nu\to\infty$. Then we consider their representation by the Mellin-Barnes integral, their asymptotic behavior and integro-differential relations. After this, in section \ref{S4}, we  generalize the obtained results to the case of operators $F = -(-\Delta)^\nu + K(\nabla) - m^2$, where $ K(\nabla)$ is an arbitrary differential operator of order less than $2\nu$ with constant coefficients. We consider general integral representations of heat kernels of these operators and discuss their asymptotic behavior. Then for arbitrary operators which are invariant with respect to $O(d)$-rotations we obtain exact analytic expressions for heat kernels and Green functions in the form of Fox--Wright psi functions and the Fox $H$-function, respectively. The results turn out to be directly applicable in the case of operators not only of integer but also of non-integer order. In Conclusion we briefly discuss the results obtained and the prospects for their development in the theory of fractional calculus.

%%%%%%%%%%%%%%%%%%%%%
\section{Proper time method} \label{S1}
%%%%%%%%%%%%%%%%%%%%%

Let $F(\nabla)$ be a differential operator in $d$-dimensional Euclidean space. Then its \emph{heat kernel} (or sometimes evolution function) $U_F(\tau; \bs{x})$ is defined \footnote{See, for example, \cite{Avram00, Avram01, Vassil03}. In the definitions used by different authors a different choice of signs occurs, as well as a rotation in the complex plane of the parameter $\tau$.} as the kernel of the operator $e^{\tau F}$, that is, as the solution of the differential equation
\begin{equation} \label{EqEvol}
\partial_\tau U_F(\tau;\bs{x}) = FU_F(\tau;\bs{x})
\end{equation}
with the initial condition
\begin{equation} \label{GranUsl}
U_F(0; \bs{x}) = \delta(\bs{x}).
\end{equation}
The parameter $\tau$ is traditionally called ``proper time''.

%%%%%%%%%%%%%%%%%%%%%
\paragraph{Integral representations of Green functions.}

The Green function of the operator $F$ is defined by the relation
\begin{equation}
FG_{F}(\bs{x}) = - \delta(\bs{x}),
\end{equation}
and can be represented as an integral of the heat kernel over its proper time.

Suppose that the operator $F$ is such that
\begin{enumerate}
\item the fundamental solution $U_F(\tau; \bs{x})$ is defined for all $\tau>0$;
\item $U_F(\tau; \bs{x}) \xrightarrow[\tau\to\infty]{} 0$ (for this it is necessary that $F$ be strictly negative and, in particular, non-degenerate);
\item as $\tau$ tends to infinity the function $U_F(\tau; \bs{x})$ decreases rapidly enough that the integrals below converge.
\end{enumerate}
It is easy to see that due to these three conditions the Green function of the operator $F-m^2$, where $m^2$ is some constant, can be represented as
\begin{equation} \label{Gm2}
G_{F-m^2}(\bs{x}) = -\frac{1}{F-m^2} = \int\limits_0^\infty d\tau e^{-m^2 \tau} U_F(\tau; \bs{x}) .
\end{equation}

The Green function of the operator $F$ raised to a natural power $\nu$ can also be expressed in terms of the heat kernel
\begin{equation} \label{FnGr}
G_{F^\nu}(\bs{x}) = -\frac{1}{F^\nu} = \frac{(-1)^{\nu-1}}{\Gamma(\nu)}\int\limits_0^\infty d\tau \tau^{\nu-1} U_F(\tau; \bs{x}).
\end{equation}
The relation (\ref{FnGr}) can be verified by alternately acting on it by the operator $F$ and integrating by parts $\nu$ times. The same answer can be obtained by differentiation the representation (\ref{Gm2}) $\nu-1$ times with respect to the parameter $m^2$, and then setting $m^2 = 0$.

It is known that the above three conditions and hence the following representations (\ref{Gm2}--\ref{FnGr}) are certainly true if $F$ is an operator of the Laplace type, obtained by adding to the Laplacian a potential term, and, accordingly, if $F^\nu$ is a minimal operator, i.e. a natural power of an operator of the Laplace type. In this paper we show that for $m^2 \ne 0 $ the conditions hold for an arbitrary $d$ at least for all operators of the form $-(-\Delta)^\nu$ with $\nu > 1/2$, and for $m^2 = 0$ for $1/2 <\nu < d/2$, and also give expressions for a wider class of operators.

%%%%%%%%%%%%%%%%%%%%%
\paragraph{The Green function of the operator $(\Delta - m^2)^\nu$.}

If $F = \Delta$ is the Laplace operator, then the equation (\ref{EqEvol}) is the heat equation
\begin{equation}
\partial_\tau U_\Delta(\tau; \bs{x}) = \Delta U_\Delta(\tau; \bs{x}),
\end{equation}
and the initial condition (\ref{GranUsl}) defines its well-known fundamental solution
\begin{equation} \label{HeatKernel}
U_\Delta(\tau;\bs{x}) = \frac{1}{(4\pi\tau)^{d/2}} \exp\left(-\frac{\bs{x}^2}{4\tau}\right).
\end{equation}

Substituting (\ref{HeatKernel}) into (\ref{FnGr}), we get the following representation of the Green function of the operator $(\Delta - m^2)^\nu$ as a proper time integral \footnote{Note that the choice of the sign ``$-$'' in front of $m^2$ is due to the fact that we are working in Euclidean space. When passing to the Minkowski space with the signature $(+ - \ldots -)$, the Laplace operator $\Delta = \sum\partial^2_i$ turns into the D'Alembert operator $-\Box = -\partial_0^2 + \partial^2_1 + \ldots + \partial_{d-1}^2$, and the operator $\Delta - m^2$ turns into the minus Klein-Gordon operator $-(\Box + m^2)$. (In the momentum representation, $-(k^2 + m^2)$ turns into $k_0^2 - \bs{k}^2 - m^2$.)}:
\begin{equation} \label{OkonInt}
G_{(\Delta-m^2)^\nu}(\bs{x}) = \frac{(-1)^{\nu-1}}{(4\pi)^{d/2} \Gamma(\nu)} \int\limits_0^\infty d\tau \tau^{\nu-\frac{d}{2}-1} \exp\left(-\frac{\bs{x}^2}{4\tau} - m^2 \tau\right).
\end{equation}

This representation can also be obtained if we write the Green function in terms of the integral in momentum space, represent the denominator $(k^2 + m^2)^\nu$ in the standard way as the integral of the exponent $e^{-(k^2 + m^2)\tau}$, isolate the complete square and take the Gaussian integral over the momenta.

The integral (\ref{OkonInt}) can be expressed in terms of Hankel functions. For them there is a well-known integral representation (see, for example, \cite[p. 21]{BE})
\begin{equation} \label{Hankel}
\pi H_\lambda^{(1)}(\alpha z) = (-i)^{\lambda+1} \alpha^\lambda \int\limits_0^\infty \exp\left\{\frac{iz}{2}\left(\tau + \frac{\alpha^2}{\tau}\right)\right\} \tau^{-\lambda-1} d\tau.
\end{equation}
It is true for $\Im z > 0$ and $\Im (z\alpha^2) > 0$. Let $\lambda = d/2 - \nu$, $z = 2im^2$, $\alpha = x/2m$ and $\alpha z = imx$, then we obtain ($x = \sqrt{\bs{x}^2}$)
\begin{equation}
G_{(\Delta-m^2)^\nu}(\bs{x}) = \frac{(-1)^{\nu-1} i\pi}{(4\pi)^{d/2}\Gamma(\nu)} \left(\frac{-ix}{2m}\right)^{\nu- \frac{d}{2}} H^{(1)}_{\frac{d}{2} - \nu}(imx).
\end{equation}
The formula can be given in a more convenient form in terms of the MacDonald functions 
\begin{equation}
G_{(\Delta-m^2)^\nu}(\bs{x}) = \frac{(-1)^{\nu-1} 2}{(4\pi)^{d/2}\Gamma(\nu)} \left(\frac{x}{2m}\right)^{\nu- \frac{d}{2}} K_{\frac{d}{2} - \nu}(mx).
\end{equation}
This expression is valid for all $\Re m^2 > 0$.

In the limit $z \gg |\alpha^2 - 1/4|$, the asymptotic behavior of the MacDonald functions has the form $K_\alpha(z) \sim \sqrt{\pi/2z} e^{-z}$. Correspondingly, for $mx \gg |(d/2 - \nu)^2 - 1/4|$
\begin{equation}
G_{(\Delta-m^2)^\nu}(\bs{x}) \sim \frac{(-1)^{\nu-1}2}{(4\pi)^{d/2}\Gamma(\nu)} \left(\frac{x}{2m}\right)^{\nu- \frac{d}{2}} \sqrt{\frac{\pi}{2mx}} e^{-mx}.
\end{equation}

In the limit $z \ll \sqrt{\alpha+1}$ we have
\begin{equation}
K_\alpha(z) \sim \begin{cases}
\frac{\Gamma(|\alpha|)}{2}\left(\frac{2}{z}\right)^{|\alpha|}, & \alpha\ne0, \\
- \ln\left(\frac{z}{2}\right) - \gamma, & \alpha=0.
\end{cases} \end{equation}
Correspondingly, in the limit $mx \ll \sqrt{|d/2 - \nu| + 1}$ we obtain the following asymptotic expression for the Green function
\begin{equation}
G_{(\Delta-m^2)^\nu}(\bs{x}) \sim \begin{cases}
C\Gamma\left(\frac{d}{2} - \nu\right) \left(\frac{4}{x^2}\right)^{\frac{d}{2} - \nu}, & d/2>\nu, \\
 C\left(\frac{x}{2m}\right)^{\nu- \frac{d}{2}}\left[-\ln\left(\frac{mx}{2}\right)-\gamma\right], & d/2=\nu, \\
C\Gamma\left(\nu-\frac{d}{2}\right) m^{d - 2\nu}, & d/2<\nu,
\end{cases}\end{equation}
where $ C = (-1)^{\nu-1} 2(4\pi)^{-d/2} / \Gamma(\nu)$. In the first case, the Green function ceases to depend on $m$, and in the third case it ceases to depend on $\bs{x}$.

Letting $m\to0$, we get that for $\nu < d/2$
\begin{equation} \label{Gnu}
G_{\Delta^\nu}(\bs{x}) = \frac{(-1)^{\nu-1} \Gamma(d/2 - \nu)}{(4\pi)^{d/2}\Gamma(\nu)} \left(\frac{4}{x^2}\right)^{\frac{d}{2} - \nu}.
\end{equation}
For $\nu \ge d/2$, the limit does not exist and the Green function $G_{\Delta^\nu}(\bs{x})$ is not defined. If we take the original integral (\ref{OkonInt}) and set $m = 0$ in it, then it will give (\ref{Gnu}) for $\nu < d/2 $, and will be divergent at large $\tau$ for $\nu \ge d/2$.

%%%%%%%%%%%%%%%%%%%%%%%%%%%%%%
\section{The heat kernel of the operator $-(-\Delta)^\nu$} \label{S2}
%%%%%%%%%%%%%%%%%%%%%%%%%%%%%%

Now we turn to the heat kernel of the operator $F = -(-\Delta)^\nu$ calculation. We denote it by $U_{\nu, d}(\tau; \bs{x})$. In this case, the equation (\ref{EqEvol}) takes the form
\begin{equation} \label{FracDiff}
\partial_\tau U_{\nu,d}(\tau;\bs{x}) = -(-\Delta)^\nu U_{\nu,d}(\tau;\bs{x}).
\end{equation}
The solution can be written in the form of the integral over the momentum space ($k = \sqrt{\bs{k}^2}$):
\begin{equation} \label{UIntP}
U_{\nu,d}(\tau;\bs{x}) = \int \frac{d^d\bs{k}}{(2\pi)^d} \exp\left(-k^{2\nu}\tau + i\bs{k}\bs{x} \right).
\end{equation}
We evaluate this integral in two different ways.

%%%%%%%%%%%%%%%%%%%%%%%%%%%%%%
\paragraph{The first way to calculate $U_{\nu,d}(\tau;\bs{x})$.}

Note that the heat kernel is invariant with respect to $O(d)$-rotations $ U_{\nu, d}(\tau; \bs{x}) = U_{\nu, d}(\tau; \sigma)$, where $\sigma = \bs{x}^2/2$, and is scale-invariant $\alpha^d U_{\nu, d}(\alpha^{2\nu}\tau; \alpha\bs{x}) = U_{\nu, d}(\tau; \bs{x})$. Therefore, it should have the form
\begin{equation} \label{FunkEvol}
U_{\nu,d}(\tau;\bs{x}) = C_0 \tau^{-\frac{d}{2\nu}} \calE_{\nu, d/2}\left(-\frac{\sigma}{2\tau^{1/\nu}}\right),
\end{equation}
where $\calE_{\nu, d/2}(z)$ is some unknown function\footnote{We use the letter $\calE$ for this function, since it stands in place of the exponent in the usual expression for the heat kernel (\ref{HeatKernel}) and in this sense can be considered as one of its possible generalizations. However, this is not a Mittag--Leffler function $E_{\alpha, \beta}(z) $, which is also considered as a generalization of the exponent and is therefore denoted by the same letter.}, and $C_0$ is the normalization constant.

Let us find the expansion of the function $\calE_{\nu, d/2}(z)$ in its Taylor series. Using the relations $\nabla\sigma^k = k \sigma^{k-1}\bs{x}$ and $\nabla(\sigma^k\bs{x}) = (d + 2k)\sigma^k$, it is easy to verify by induction that for an arbitrary function $f(\alpha\sigma)$, where $\alpha$ is a constant, the following formula holds
\begin{equation}
\Delta^m f(\alpha\sigma) = (2\alpha)^m \sum\limits_{k=0}^m C_m^k \frac{\Gamma(d/2+m)}{\Gamma(d/2+k)}(\alpha\sigma)^k f^{(m+k)}(\alpha\sigma).
\end{equation}

Setting $f = \calE_{\nu,d/2}$, $\alpha=-1/2\tau^{1/\nu}$, $\sigma=0$, we obtain
\begin{equation} \label{Proizv2}
(-\Delta)^m U_{\nu,d}(\tau; 0) = C_0 \tau^{-\frac{d/2+m}{\nu}} \frac{\Gamma(d/2+m)}{\Gamma(d/2)} \calE_{\nu,d/2}^{(m)}(0).
\end{equation}

On the other hand, these quantities can easily be calculated directly
\begin{multline} \label{Proizv1}
(-\Delta)^m U_{\nu,d}(\tau;0) = \int \frac{d^d\bs{k}}{(2\pi)^d} k^{2m} \exp \left(- k^{2\nu}\tau \right) = \\
\frac{1}{(2\pi)^d} \frac{2\pi^{d/2}}{\Gamma(d/2)} \int\limits_0^\infty dk k^{2m+d-1} \exp(-k^{2\nu}\tau) =
\frac{\tau^{-\frac{d/2+m}{\nu}} \Gamma\left(\frac{d/2+m}{\nu}\right)}{(4\pi)^{d/2}\nu\Gamma(d/2)}.
\end{multline}

Comparing the expressions (\ref{Proizv2}) and (\ref{Proizv1}), we find that for the choice of the normalization $C_0 = 1/(4\pi)^{d/2}$, the function $\calE_{\nu, \alpha}(z) $ is determined by the following Taylor series \footnote{In \cite{Wach} we did not include the factor $1/\nu$ in the definition of the function $\calE_{\nu, \alpha}(z)$. The change of the definition is connected, in particular, with the fact that in the new notation it is more convenient to consider the limit $\nu\to\infty$ (\ref{NuInfty}).}
\begin{equation} \label{Em}
\calE_{\nu,\alpha}(z) = \frac{1}{\nu}\sum\limits_{m=0}^\infty \frac{\Gamma\left(\frac{\alpha+m}{\nu}\right) }{\Gamma(\alpha+m)} \frac{z^m}{m!}.
\end{equation}

%%%%%%%%%%%%%%%%%%%%%%%%%%%%%%
\paragraph{The second way to calculate $U_{\nu,d}(\tau;\bs{x})$.}

The expression (\ref{Em}) can be obtained in another way. Let the angle between the vectors $\bs{x}$ and $\bs{p}$ in the expression (\ref{UIntP}) is $\theta$. Integrating over all other angles, we obtain
\begin{equation} \label{IntUgol}
U_{\nu,d}(\tau;\bs{x}) = \frac{1}{(2\pi)^d} \frac{2\pi^{(d-1)/2}}{\Gamma(\frac{d-1}{2})} \int\limits_0^\infty e^{-k^{2\nu}\tau} k^{d-1}dk \int\limits_0^\pi e^{ikx\cos\theta} \sin^{d-2}\theta d\theta.
\end{equation}

We expand $\exp(ikx\cos\theta)$ in a Taylor series and integrate over $\theta$, using the fact that the integral $\int_0^\pi \cos^m\theta \sin^n\theta d\theta$ is equal to zero for odd $m$ and $B(\frac{m+1}{2}, \frac{n+1}{2})$ for even $m$, then
\begin{multline} \label{ThetaInt}
\int\limits_0^\pi \exp(ikx\cos\theta) \sin^{d-2}\theta d\theta = \sum\limits_{m=0}^\infty \frac{(ikx)^m}{m!} \int\limits_0^\pi \cos^m\theta \sin^{d-2}\theta d\theta = \\
\sum\limits_{m=0}^\infty \frac{(ikx)^{2m}}{(2m)!} B\left(\frac{d-1}{2}, m+\frac{1}{2}\right) = \sqrt{\pi}\Gamma\left(\frac{d-1}{2}\right) \sum\limits_{m=0}^\infty \frac{(-x^2)^m k^{2m}}{4^m m!\Gamma\left( \frac{d}{2}+m \right)}.
\end{multline}
Here we used the Legendre duplication formula $\sqrt{\pi}\Gamma(2m+1) = 2^{2m}\Gamma(m+1/2)\Gamma (m+1)$.

Substituting this result in (\ref{IntUgol}) and integrating over $k$ (this integral is exactly the same as in the formula (\ref{Proizv1})), we obtain
\begin{multline} \label{k2Int}
U_{\nu,d}(\tau;\bs{x}) = \frac{2}{(4\pi)^{d/2}} \sum\limits_{m=0}^\infty \frac{(-x^2/4)^m}{m!\Gamma(d/2+m)} \int\limits_0^\infty \exp(-k^{2\nu}\tau) k^{2m+d-1}dk = \\
\frac{\tau^{-d/2\nu}}{(4\pi)^{d/2}\nu} \sum\limits_{m=0}^\infty \frac{\Gamma\left(\frac{d/2+m}{\nu}\right)}{m!\Gamma(d/2+m)} \left(\frac{-x^2}{4\tau^{1/\nu}}\right)^m.
\end{multline}
The resulting expression coincides with the result given by the formulae (\ref{FunkEvol}) and (\ref{Em}).

\begin{figure} 
	\begin{minipage}[b]{0.5\linewidth}
		\includegraphics[scale=0.6]{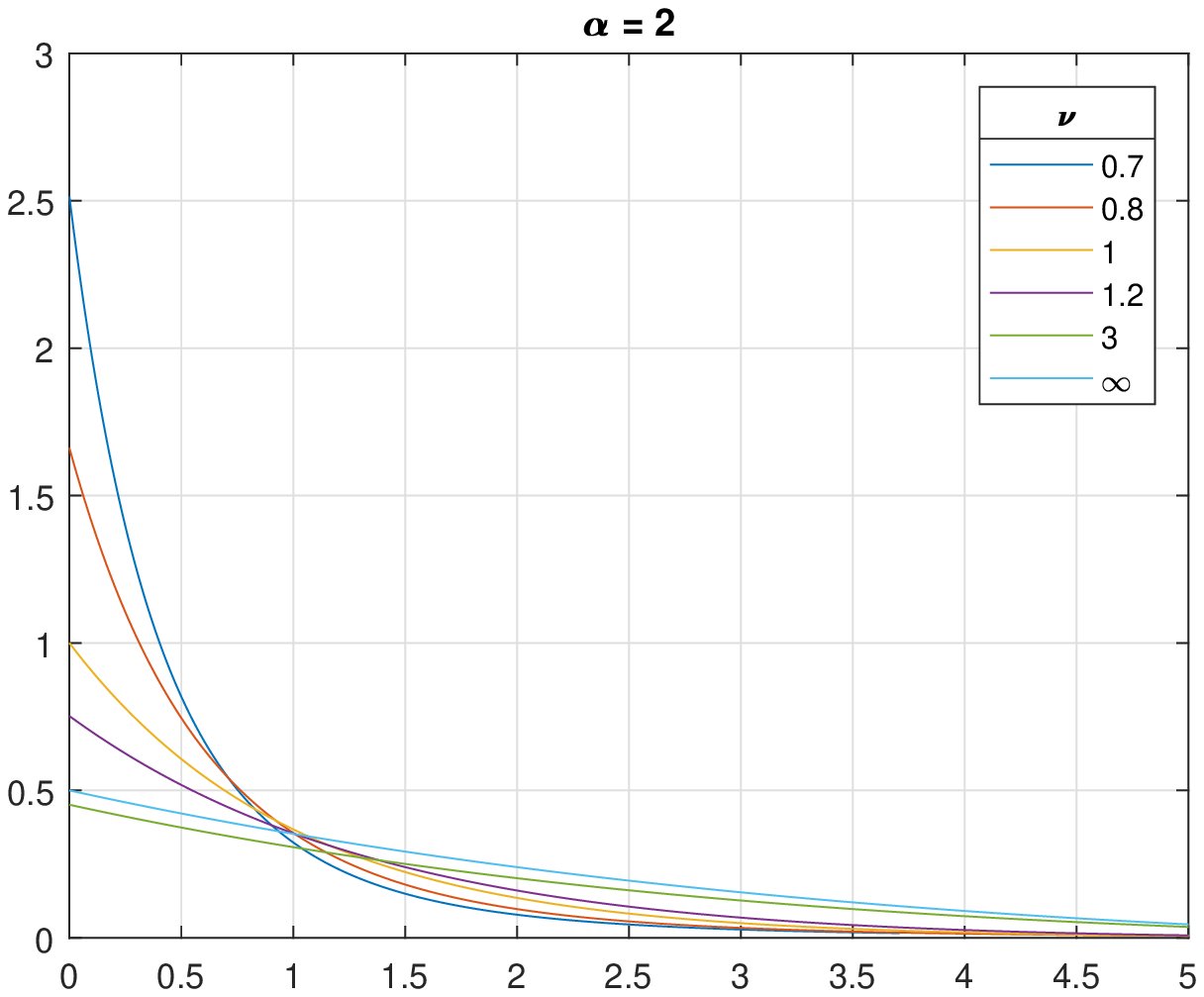}
	\end{minipage}
	\begin{minipage}[b]{0.5\linewidth}
		\includegraphics[scale=0.6]{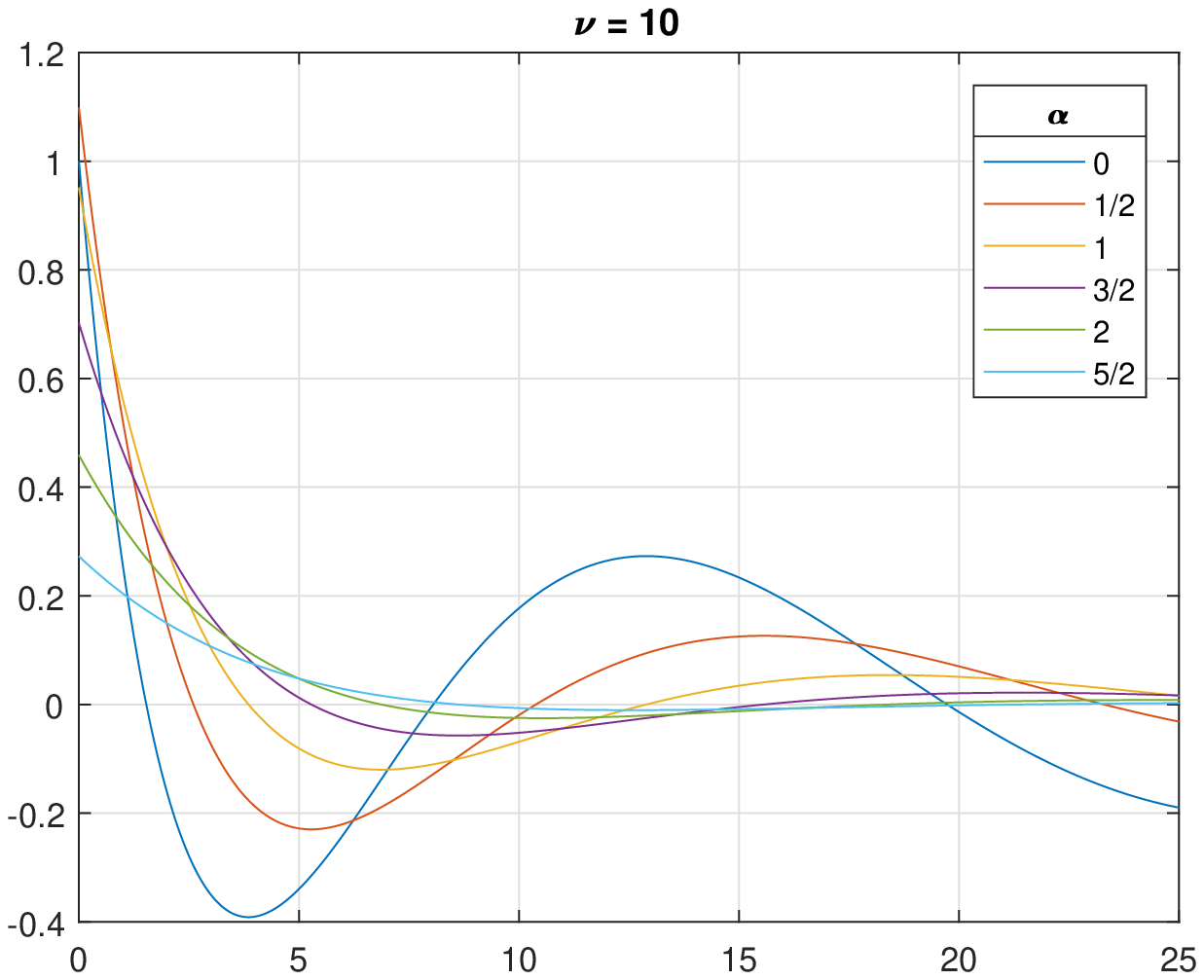}
	\end{minipage}
\begin{center} \caption{Graphs of the function $\calE_{\nu, \alpha}(-z)$ for various values of the parameters $\nu$ and $\alpha$. For $\nu = \infty$, the graph of the limit function $\calC_\alpha(-z)$ is given (\ref{NuInfty}).} \label{Fig1} \end{center}
\end{figure}

\begin{figure} 
	\begin{minipage}[b]{0.5\linewidth}
		\includegraphics[scale=0.6]{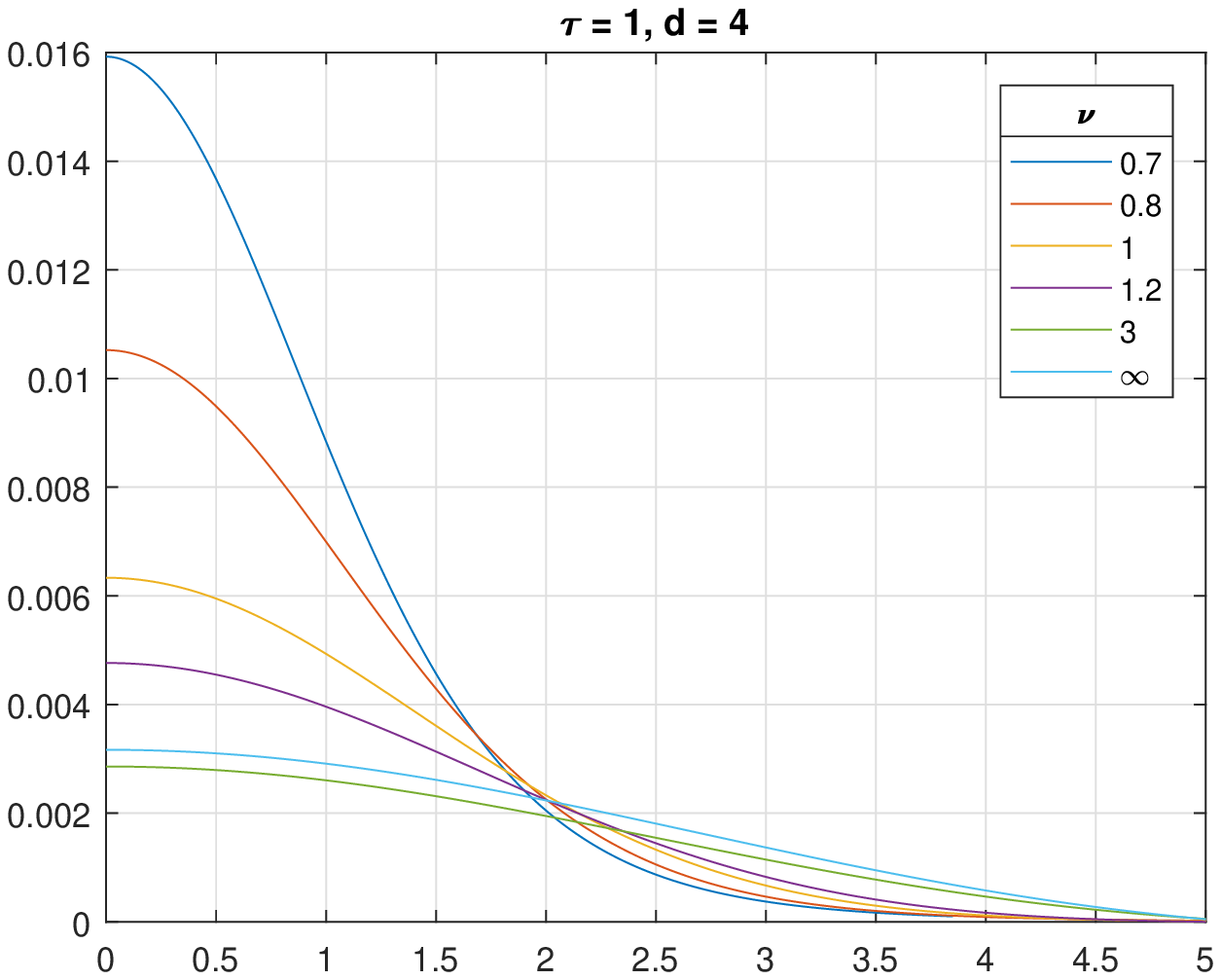}
	\end{minipage}
	\begin{minipage}[b]{0.5\linewidth}
		\includegraphics[scale=0.6]{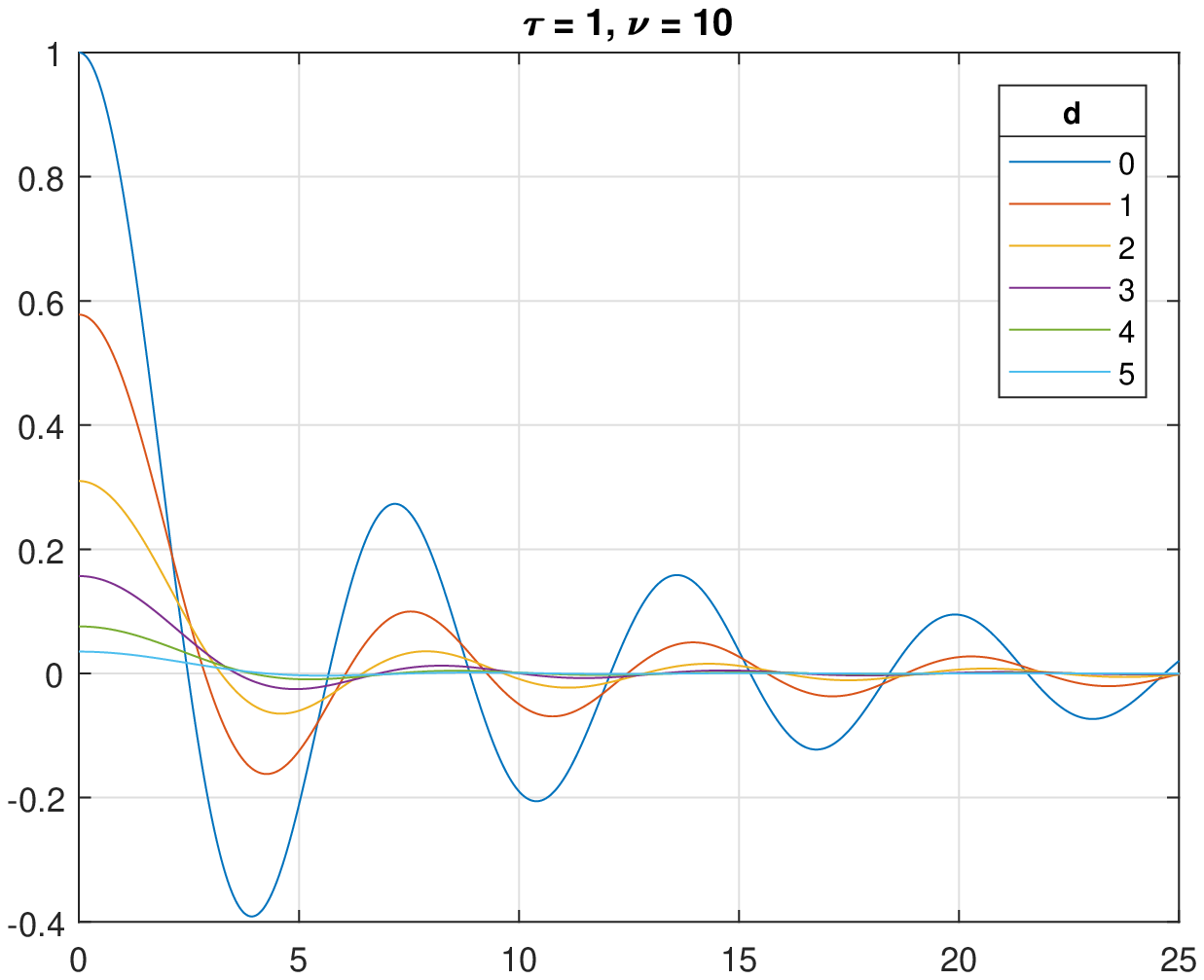}
	\end{minipage}
\begin{center}\caption{Graphs of the function $U_{\nu,d}(1, x)$ for various values of the parameters $\nu$ and $\alpha$. For $\nu = \infty$, the graph of the limit function $U_{\infty, d}(x)$ is given (\ref{UInfty}).} \label{Fig2} \end{center}
\end{figure}

%%%%%%%%%%%%%%%%%%%%%%%%%%%%%%%%%%
\section{$\calE_{\nu,\alpha}(z)$ functions} \label{S3}
%%%%%%%%%%%%%%%%%%%%%%%%%%%%%%%%%%

%%%%%%%%%%%%%%%%%%%%%%%%%%%%%%%%%%
\paragraph{Representation via Fox--Wright psi functions.}

It follows from the expansion (\ref{Em}) that
\begin{equation} \label{EPsi}
\calE_{\nu,\alpha}(z) = \frac{1}{\nu} {}_1\Psi_1\left[ \left(\frac{\alpha}{\nu},\frac{1}{\nu}\right); (\alpha, 1); z \right],
\end{equation}
where ${}_p\Psi_q[(a,A); (b,B); z]$ are the Fox--Wright psi functions defined by their Taylor series
\begin{equation} \label{PsiFunction}
{}_p\Psi_q[(a,A); (b,B); z] = \sum\limits_{k=0}^\infty \frac{\Gamma(a_1+A_1k)\ldots\Gamma(a_p+A_pk)}{\Gamma(b_1+B_1k)\ldots\Gamma(b_q+B_qk)} \frac{z^k}{k!}.
\end{equation}
These special functions are one of the possible further extensions of the generalized hypergeometric series (${}_pF_q[a; b; z] = {}_p\Psi_q[(a,1); (b,1); z]\Gamma(b)/\Gamma(a)$) and have applications, in particular, in fractional calculus \cite{SKM, Pschu, Kilbas05, Lav17, Mainardi96, Gorenflo00, Mainardi01}. They were introduced by E.~M.~Wright, who studied their asymptotic behavior \cite{Wright35, Wright40}. In recent years, the properties of the Fox--Wright psi functions have been investigated in detail in the papers \cite{Kilbas02, Kilbas06, Kilbas09, Meh17}.

Thus, we have got the following representation of the heat kernel
\begin{equation} \label{FunkEvolOkon}
U_{\nu,d}(\tau;\bs{x}) = \frac{\tau^{-d/2\nu}}{(4\pi)^{d/2}\nu} {}_1\Psi_1\left[ \left(\frac{d}{2\nu},\frac{1}{\nu}\right); \left(\frac{d}{2}, 1\right); -\frac{\bs{x}^2}{4\tau^{1/\nu}} \right].
\end{equation}

The expression (\ref{FunkEvolOkon}) is a generalization of the well-known heat kernel. Indeed, for $\nu=1$
\begin{equation}
\calE_{1,\alpha}(z) = {}_1\Psi_1[(\alpha,1); (\alpha,1); z] = e^z.
\end{equation}
Substitution this expression in (\ref{FunkEvolOkon}) gives us the result of (\ref{HeatKernel}).

%%%%%%%%%%%%%%%%%%%%%%%%%%%%%%%%%%
\paragraph{The functions' domain.}

The terms of (\ref{PsiFunction}) are well defined for the parameters $(a_1, A_1)$, \dots, $(a_p, A_p)$ such that $a_j + A_j m \ne 0, -1, -2, \ldots$ for all $j = 1, \ldots, p$ and all $m$. The series converges absolutely on the whole complex plane $z$ if $\delta = \sum_{j=1}^q B_j - \sum_{j=1}^p A_j > -1$ for positive $A_j$ and $B_j$.

As applied to the function $\calE_{\nu, \alpha}(z)$, both these conditions are satisfied for all real $\nu>1/2$ and complex
\begin{equation} \label{UslOpr}
\alpha\ne-m-n\nu, \quad \text{where $m,n = 0, 1, 2, \ldots$}
\end{equation}

We note, however, that the singularities at the points $\alpha = -m$ (and hence also at the points $\alpha = -n\nu$ for integers $\nu$) are removable, since the poles of the gamma functions in the numerator and denominator cancel each other. Expanding $\Gamma(-n+z) \sim (-1)^n/z^{n+1}$ it is possible to define
\begin{gather}
\calE_{\nu,0}(z) = 1 + \frac{1}{\nu}\sum\limits_{m=1}^\infty \frac{\Gamma\left(\frac{m}{\nu}\right) }{\Gamma(m)} \frac{z^m}{m!}, \\
\calE_{\nu,-1}(z) = z + \frac{1}{\nu}\sum\limits_{m=2}^\infty \frac{\Gamma\left(\frac{m-1}{\nu}\right) }{\Gamma(m-1)} \frac{z^m}{m!} \quad\text{and so on.}
\end{gather}
Non-removable poles remain only for $\alpha = -n\nu$ for non-integer $\nu$.

Thus, for natural $\nu$ the function $\calE_{\nu, \alpha}(z)$ is an entire function of $z$ for any values of $\alpha$, and for non-integer $\nu>1/2$, for $\alpha \ne -n\nu$. Consequently, the functions $U_{\nu, d}(\tau; \bs{x})$ are well defined not only for all natural numbers, but also for fractional $\nu$  and $d$ satisfying these conditions.

\begin{figure} \begin{center}
	\includegraphics[scale=0.8]{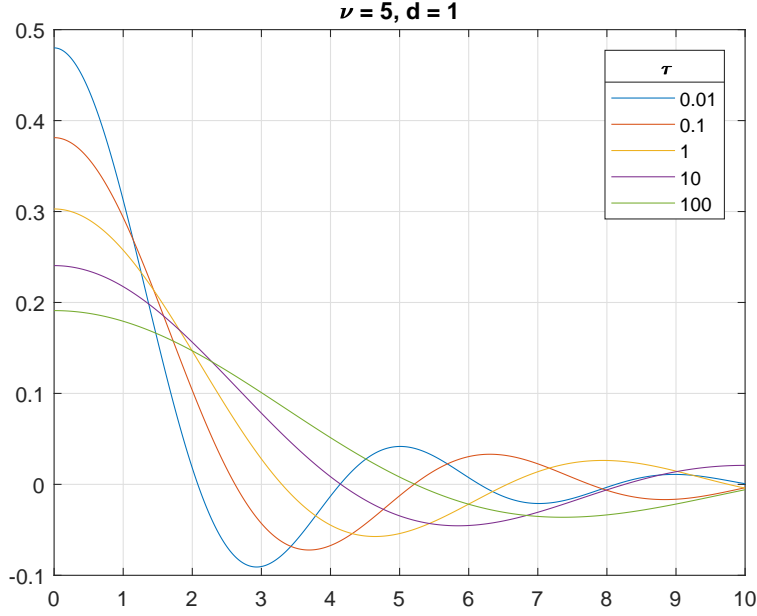}
	\caption{Graphs of the function $U_{5,1}(\tau, x)$ for different values of the proper time $\tau$.} \label{Fig3} \end{center}
\end{figure}

The graphs of the functions $\calE_{\nu, \alpha}(z)$ and $U_{\nu, d}(\tau; x)$ for various values of the parameters, obtained by numerical summation of the series (\ref{Em}) in MATLAB, are shown in Fig. \ref{Fig1}--\ref{Fig3}. One can see that one of the main features of these functions is that they oscillate for $\nu \ne 1$. When approaching the critical value $\nu = 1/2$, the series (\ref{Em}) begin to converge very badly and are difficult to evaluate.

%%%%%%%%%%%%%%%%%%%%%%%%%%%%%%%%%%
\paragraph{Connection with the operators of fractional integro-differentiation.}

For non-integer $\nu$ the resulting expression (\ref{FunkEvolOkon}) is the solution of the equation (\ref{FracDiff}), in which the operator $-(-\Delta)^\nu $ should be understood as the so-called fractional Riesz derivative of order $2\nu$, defined using the Fourier transform \cite{SKM}:
\begin{equation}
D^\alpha_{Riesz} f(\bs{x}) = \calF^{-1}(k^\alpha f(\bs{k})), \quad f(\bs{k}) = \calF(f(\bs{x})).
\end{equation}
The corresponding equations are called fractional diffusion equations and have been widely discussed in the mathematical literature, see, for example, \cite{Pschu, ZUS, Mainardi96, Gorenflo00, Mainardi01}. However, in such  papers fractional equations in $(1+1)$-dimensional space are usually considered, i.e. the case $d=1$ in our notation.

We note that a series analogous to the series (\ref{k2Int}) was given in \cite{ZUS} in the context of the study of anomalous diffusion in $d$-dimensional space and spherically symmetric stable distributions. However, in this paper the parameter $\nu$ is bounded by the interval $(1/2, 1)$ and no expressions are given in terms of the Fox--Wright psi functions.

%%%%%%%%%%%%%%%%%%%%%%%%%%%%%%%%%%
\paragraph{The $\nu\to\infty$ limit.}

Note that the functions $\calE_{\nu, \alpha}$ have a limit at $\nu\to\infty$. Indeed, replacing $\Gamma(c/\nu) \sim \nu/c$ in the expansion (\ref{Em}), we obtain
\begin{gather}
\calE_{\nu,\alpha}(z) \xrightarrow[\nu\to\infty]{} \calE_{\infty,\alpha}(z) = \calC_\alpha(z), \label{NuInfty}\\
\text{where}\quad \calC_\alpha(z) = \frac{1}{\Gamma(\alpha+1)} {}_0F_1(\alpha+1; z) = \sum\limits_{m=0}^\infty \frac{1}{\Gamma(\alpha+1+m)}\frac{z^m}{m!}
\end{gather}
are Bessel--Clifford functions.

Using the well-known connection between the Bessel--Clifford and Bessel functions
\begin{equation}
J_\alpha(x) = \left(\frac{x}{2}\right)^\alpha \calC_\alpha\left(-\frac{x^2}{4}\right),
\end{equation}
we obtain the limit of the functions $U_{\nu,d}(\tau; x)$ at $\nu\to\infty$
\begin{equation} \label{UInfty}
U_{\infty,d}(x) = (4\pi)^{-d/2}\calC_{d/2}\left(-\frac{x^2}{4}\right) = (2\pi x)^{-d/2} J_{d/2}(x).
\end{equation}

%%%%%%%%%%%%%%%%%%%%%%%%%%%%%%%%%%%%%%
\paragraph{Representation via the Mellin--Barnes integral.}

The Fox--Wright psi functions are a special case of more general Fox $H$-functions. The latter are defined in terms of the Mellin--Barnes integral
\begin{equation}
H_{p,q}^{m,n}\left[z \left|\begin{smallmatrix}(a,A)\\ (b, B)\end{smallmatrix}\right.\right] =
\frac{1}{2\pi i} \int\limits_C \frac{\prod\limits_{i=1}^m \Gamma(b_i-B_is) \prod\limits_{j=1}^n \Gamma(1-a_j+A_js)}{\prod\limits_{j=m+1}^q \Gamma(1-b_i+B_is) \prod\limits_{j=n+1}^p \Gamma(a_j-A_js)} z^s ds,
\end{equation}
where the contour of integration $C$ is chosen to pass through infinity and to separate the poles of $\Gamma(b_i-B_is)$ and $\Gamma(1-a_j + A_js)$. The Fox $H$-functions are in exactly the same way related to Fox--Wright psi functions, as the well-known Meyer $G$-functions to generalized hypergeometric functions. The general theory of $H$-functions and $H$-transforms can be found in \cite{Braaksma, Sri84, MSH, KS}. Psi functions are expressed in terms of the $H$-function
\begin{equation}
{}_p\Psi_q\left[\left.\begin{smallmatrix}(a,A)\\ (b, B)\end{smallmatrix}\right| z\right] = H_{p,q+1}^{1,p}\left[-z \left|\begin{smallmatrix}(1-a,A)\\ (0,1), (1-b, B)\end{smallmatrix}\right.\right].
\end{equation}

In our case, these general results allow us to obtain the following representation of the function $\calE_{\nu, \alpha}(z)$ by the Mellin--Barnes integral
\begin{equation} \label{MellinBurns}
\calE_{\nu,\alpha}(z) = \frac{1}{2\pi i} \int\limits_C \frac{\Gamma(-s)\Gamma\left(\frac{\alpha+s}{\nu}\right)}{\nu\Gamma(\alpha+s)} (-z)^s ds.
\end{equation}
It is not difficult to see that this integral actually gives the required decomposition (\ref{Em}). Indeed, the poles of $\Gamma(-s)$ lie at the points $s_m = m$, and the poles of $\Gamma\left((\alpha + s)/\nu\right)$ lie at the points $s_k = -\alpha - k\nu$. The condition (\ref{UslOpr}) simply means that they do not merge, and we can separate them with some contour $C$. For $\nu>1/2$, we can close this contour on the right so that the imaginary part of $s$ remains bounded. Then the integral is equal to the sum of the residues at the poles of $\Gamma(-s)$.

 %%%%%%%%%%%%%%%%%%%%%%%%%%%%%%%%%%%%%%
\paragraph{The asymptotic behavior of $\calE_{\nu,\alpha}(z)$.}

It is convenient to use the representation (\ref{MellinBurns}) to study various properties of the functions $\calE_{\nu, \alpha}(z)$. Thus, the power-law part of their asymptotic expression can be obtained if we formally close the contour $C$ on the left and sum the residues at the poles of $\Gamma((\alpha + s)/\nu)$.

However, not all the poles $s_k = -\alpha - k\nu$ of the function $\Gamma((\alpha + s)/\nu)$ in the numerator will be the poles of the entire integrand, since some of them will be canceled by the poles $s_j = -\alpha-j$ of the function $\Gamma(\alpha+s)$ in the denominator. So, $s_0 = -\alpha$ is never a pole. If $\nu = p/q$ is an irreducible fraction, then $s_q, s_{2q}, s_{3q}, \ldots$ will also not be poles. Finally, if $\nu$ is a natural number, then all poles are eliminated without exception.

Suppose, however, that $\nu$ is not an integer. Then the sum over all non-canceling poles gives the following asymptotic expansion
\begin{equation} \label{Asympt}
\calE_{\nu,\alpha}(-z) \sim -z^{-\alpha} \sum\limits_m \frac{(-1)^m}{m!}\frac{\Gamma(\alpha+m\nu)}{\Gamma(-m\nu)} z^{-m\nu}.
\end{equation}
The term with $m = 1$ is the leading term of the asymptotic expression as $z\to \infty$
\begin{equation} \label{Asympt2}
\calE_{\nu,\alpha}(-z) \sim \frac{\Gamma(\alpha+\nu)}{\Gamma(-\nu)} z^{-\alpha-\nu}.
\end{equation}

At the same time, for a natural $\nu$, all terms in the expansion (\ref{Asympt}) vanish. This means that in this case the function $\calE_{\nu, \alpha}(-z)$ decreases faster than any power of $z$, i.e. in an exponential manner.

This behavior is due to the fact that ordinary derivatives of integer order $\nu$ are local operations, i.e. their value at each point is determined only by the values of the differentiable function in a small neighborhood of this point. Conversely, fractional integro-differentiation operators corresponding to non-integer values of $\nu$ are essentially non-local operations, since the value of the fractional derivative of a function at each point depends on the behavior of this function on the whole domain of its definition.

%%%%%%%%%%%%%%%%%%%%%%%%%%%%%%%%%%%%%%
\paragraph{Integro-differential relations.}

The functions $\calE_{\nu, \alpha}(z)$ are characterized by the following simple differentiation relation
\begin{equation}
\frac{d^\beta}{dz^\beta} \calE_{\nu,\alpha}(z) = \calE_{\nu,\alpha+\beta}(z).
\end{equation}
For natural $\beta$, it can be verified directly by differentiating the definition (\ref{EPsi}).

However, this relation can make sense not only for natural $\beta$, but also for all such $\beta$ that $\alpha + \beta \ne -n\nu$ (for non-integer $\nu$, and for any complex $\beta$ for natural $\nu$). For negative integers $\beta$, it will give the principal primitive of the function $\calE_{\nu, \alpha}(z)$ \footnote{This may raise the question of what happens when $\alpha + \beta = -n\nu$. Of course, primitive functions always exist, but in this case none of them would belong to the family $ \calE_{\nu, \lambda}(z)$}. For non-integer $\beta$, the symbol $d^\beta/dz^\beta$ should be understood as a certain operator of fractional integro-differentiation.

We note that for each $1/2 < \nu \le \infty$ the family of functions $\calE_{\nu, \alpha}$ is closed under the operation of differentiation. In this respect they differ favorably from the Mittag--Leffler functions $E_{\alpha, \beta}(z) = {}_1\Psi_1[(0,1); (\alpha, \beta); z]$, which set is not closed.

The representation (\ref{MellinBurns}) of the function $\calE_{\nu, \alpha}(z)$ by the Mellin--Barnes integral is an inverse Mellin transform. Therefore, the direct Mellin transform has the form ($s>0$)
\begin{equation} \label{Mellin}
\int\limits_0^\infty z^{s-1} \calE_{\nu,\alpha}(-z) dz = \frac{\Gamma\left(\frac{\alpha-s}{\nu}\right)\Gamma(s)}{\nu\Gamma(\alpha-s)}.
\end{equation}
This integral relation allows simple integration of various expressions containing functions $\calE_{\nu, \alpha}(z)$, and therefore is extremely important for our further research.

%%%%%%%%%%%%%%%%%%%%%%%%%%%%%%%%%%%%%%
\paragraph{The Green function $G_{-(-\Delta)^\nu}(\bs{x})$.}

As an example, we use the expression (\ref{Mellin}) to integrate the heat kernel over the proper time and obtain the Green function of the operator $-(-\Delta)^\nu$
\begin{multline} \label{GrinNu}
G_{-(-\Delta)^\nu}(\bs{x}) = \int\limits_0^\infty U_{\nu,d}(\tau;\bs{x})d\tau = \frac{1}{(4\pi)^{d/2}} \int\limits_0^\infty \tau^{d/2\nu} \calE_{\nu,d/2}\left(-\frac{x^2}{4\tau^{1/\nu}}\right) d\tau = \\
\frac{\nu}{(4\pi)^{d/2}} \left(\frac{4}{x^2}\right)^{\frac{d}{2}-\nu} \int\limits_0^\infty \mu^{\frac{d}{2}-\nu-1} \calE_{\nu,d/2}(-\mu) d\mu = \frac{\Gamma(\frac{d}{2}-\nu)}{(4\pi)^{d/2}\Gamma(\nu)} \left(\frac{4}{x^2}\right)^{\frac{d}{2}-\nu}.
\end{multline}
This result for $d/2>\nu$ coincides with the expression (\ref{Gnu}), which we obtained using the standard heat kernel method. For $\nu>d/2$, the integral diverges for large $\tau$ and $G_{- (-\Delta)^\nu}(\bs{x})$ does not exist.

%%%%%%%%%%%%%%%%%%%%%%%%%%%%%%%%%%
\section{Operators of the form $F = -(-\Delta)^\nu + K(\nabla) - m^2$} \label{S4}
%%%%%%%%%%%%%%%%%%%%%%%%%%%%%%%%%%

%%%%%%%%%%%%%%%%%%%%%%%%%%%%
\paragraph{General integral representations.}

Now we consider a generalization of the results obtained for operators of a more complicated form
\begin{gather}
F = -(-\Delta)^\nu + K(\nabla) - m^2, \label{ConstCoeff} \\
\text{where}\quad K(\nabla) = \sum\limits_{m=1}^{2\nu-1} K^{\lambda_1\ldots\lambda_m}\nabla_{\lambda_1}\ldots\nabla_{\lambda_m}
\end{gather}
is an arbitrary operator of order less than $2\nu$ with constant numerical coefficients $K^{\lambda_1 \ldots \lambda_m}$. Its heat kernel can be written in the form of the following integral over the momentum space
\begin{equation} \label{UFObsch}
U_F(\tau;\bs{x}) = \int \frac{d^d\bs{k}}{(2\pi)^d} \exp\left\{[- k^{2\nu} + K(\bs{k}) - m^2]\tau + i\bs{k}\bs{x} \right\},
\end{equation}
where $K(\bs{k})$ is a polynomial in $\bs{k}$ of degree at most $2\nu-1$ obtained from $K(\nabla)$ by the formal substitution $\nabla \to i\bs{k}$.

Using the scale transformation $\bs{k} \to \tau^{-1/2\nu}\bs{k}$, we reduce the integral to the form
\begin{equation} \label{UFNorm}
U_F(\tau;\bs{x}) = e^{-m^2 \tau} \tau^{-d/2\nu} \int \frac{d^d\bs{k}}{(2\pi)^d} e^{- k^{2\nu} +  i\bs{k}\bs{y}} \exp\left\{ \tau K\left(\frac{\bs{k}}{\tau^{1/2\nu}}\right)\right\},
\end{equation}
where $\bs{y} = \bs{x}/\tau^{1/2\nu}$. Then we expand the exponent $e^{\tau K}$ in a power series in $\tau^{1/2\nu}$
\begin{equation} \label{ExpK}
\exp\left\{\tau K\left(\frac{\bs{k}}{\tau^{1/2\nu}}\right)\right\} = \sum\limits_{j=0}^\infty \tau^{j/2\nu} b_j(\bs{k}),
\end{equation}
where $b_j(\bs{k})$ are polynomials in $\bs{k}$ of degree at most $(2\nu-1)j$ whose coefficients are uniquely determined by the coefficients of the polynomial $K(\bs{ k})$, $b_0 = 1$.

Substituting the expansion (\ref{ExpK}) into (\ref{UFNorm}), we obtain
\begin{gather}
U_F(\tau;\bs{x}) = e^{-m^2 \tau} \frac{\tau^{- d/2\nu}}{(4\pi)^{d/2}} \sum\limits_{j=0}^\infty \tau^{j/2\nu} B_j\left(\frac{\bs{x}}{\tau^{1/2\nu}}\right), \\
\text{where}\quad B_j(\bs{y}) = \frac{1}{\pi^{d/2}} \int d^d\bs{k} b_j(\bs{k}) e^{- k^{2\nu} +  i\bs{k}\bs{y}}. \label{IntB}
\end{gather}
The functions $B_j(\bs{y})$ are bounded and tend to zero as $\bs{y}\to\infty$. Since $B_0(\bs{y}) = \calE_{\nu, d/2}(-\bs{y}^2/4)$, we have finally
\begin{gather}
U_F(\tau;\bs{x}) = U_{\nu,d}(\tau;\bs{x}) e^{-m^2 \tau} B_K\left(\tau^{1/2\nu}; \frac{\bs{x}}{\tau^{1/2\nu}}\right), \\
\text{where}\quad B_K(\tau^{1/2\nu}; \bs{y}) = 1 + B_0^{-1}(\bs{y})\sum\limits_{j=1}^\infty \tau^{j/2\nu} B_j(\bs{y}).
\end{gather}

The function $B_K(\tau^{1/2\nu}; \bs{y}) $ is the correction to the ``unperturbed'' heat kernel $U_{\nu, d}(\tau; \bs{x})$ due to the introduction into the operator $F$ the term $K(\nabla)$ with lower order derivatives. It is easy to see that $B_K \to 1$ for $\tau \to 0 $ and, consequently, $U_F \sim U_{\nu, d}$, as it should be, because for small values of $\tau$ the behavior of the heat kernel is determined by a term with higher derivatives $-(-\Delta)^\nu$ and should not depend on $K(\nabla)$.

Note, however, that this correction differs significantly from the function $\Omega(\tau; x, x')$ in the standard heat kernel method: the latter is analytic in $\tau$, while the expansion of $B_K(\tau^{1/2\nu}; \bs{x}/\tau^{1/2\nu})$ contains both positive and negative powers of $\tau^{1/2\nu}$. This difference is due to qualitatively different types of ``perturbations'' in these two cases: we introduced new derivatives into the operator, leaving the space flat. In the method of the heat kernel, first-order derivatives that can be added to the minimal operator can be eliminated by redefining the connection. Therefore, the perturbation reduces to the appearance of nonzero curvature (that is, nonzero commutators of covariant derivatives) and of a potential term depending on the point.

%%%%%%%%%%%%%%%%%%%%%%%%%%%%
\paragraph{$O(d)$-invariant operators.}

We consider operators of the form
\begin{equation} \label{SphSymm}
K(-\Delta) = \sum\limits_{j=1}^{N} \gamma_j (-\Delta)^{\varkappa_j}, \qquad K(k^2) = \sum\limits_{j=1}^{N} \gamma_j k^{2\varkappa_j},
\end{equation}
where the integer degrees $\varkappa_j<\nu$. In this case, we can take the integrals (\ref{IntB}) in a manner analogous to the second way of computing the function $U_{\nu, d}(\tau, \bs{x})$ by expressing $B_j(\bs{y})$ in terms of the Fox--Wright psi functions.

To do this, we introduce multi-indices
\begin{equation}
\gamma = (\gamma_1,\ldots,\gamma_N),\quad \varkappa = (\varkappa_1,\ldots, \varkappa_N),\quad n = (n_1, \ldots, n_N).
\end{equation}
We denote, in the standard way,
\begin{gather}
|n| = n_1+\ldots+n_N,\qquad n\varkappa = n_1\varkappa_1+\ldots+n_N\varkappa_N, \\
\gamma^n = \gamma_1^{n_1}\ldots \gamma_N^{n_N},\qquad n! = n_1!\ldots n_N!
\end{gather}

In this notation, the expansion of the exponent will be
\begin{gather}
\exp\left\{\tau K\left(\frac{k^2}{\tau^{1/\nu}}\right) \right\} = \sum\limits_n \frac{\gamma^n}{n!} \tau^{|n|-n\varkappa/\nu} k^{2n\varkappa} = \sum\limits_{j=0}^\infty b_{2j}(k^2) \tau^{j/\nu}, \\
\text{where}\quad b_{2j}(k^2) = \sum\limits_{|n|\nu-n\varkappa=j} \frac{\gamma^n}{n!} k^{2n\varkappa}
\end{gather}
are polynomials in $k^2$ of degree at most $(\nu-1)j$.

Integrating over the angles and $k^2$ as in (\ref{ThetaInt}), we get
\begin{multline}
B_{2j}\left(-\frac{y^2}{4}\right) = 2 \sum\limits_{m=0}^\infty \frac{(-y^2/4)^m}{m!\Gamma\left(d/2 + m\right)} \int\limits_0^\infty dk k^{d+2m-1} b_{2j}(k^2) e^{-k^{2\nu}} = \\
\frac{1}{\nu}\sum\limits_{|n|\nu-n\varkappa=j} \frac{\gamma^n}{n!} \sum\limits_{m=0}^\infty \frac{\Gamma\left(\frac{d/2+n\varkappa+m}{\nu} \right)}{m!\Gamma\left(d/2 + m\right)} \left(-\frac{y^2}{4}\right)^m = \sum\limits_{|n|\nu-n\varkappa=j} \frac{\gamma^n}{n!} \calE_{\nu,d/2}^{n\varkappa}\left(-\frac{y^2}{4}\right),
\end{multline}
where we have introduced the notation
\begin{equation}
\calE_{\nu,\alpha}^\beta(z) = \frac{1}{\nu} {}_1\Psi_1\left[\left(\frac{\alpha+\beta}{\nu}, \frac{1}{\nu}\right); (\alpha, 1); z \right].
\end{equation}

Finally, for the heat kernel, we have
\begin{gather}
U_F(\tau; \bs{x}) = \frac{e^{-m^2 \tau} \tau^{-d/2\nu}}{(4\pi)^{d/2}} \sum\limits_{j=0}^\infty a_j(\tau^{1/\nu}) \calE_{\nu,d/2}^j \left( -\frac{x^2}{4\tau^{1/\nu}} \right), \label{IntKappa} \\
\text{where}\quad a_j(\tau^{1/\nu}) = \sum\limits_{n\varkappa=j} \frac{\gamma^n}{n!} \tau^{|n|-n\varkappa/\nu}
\end{gather}
are polynomials in $\tau^{1/\nu}$ of degree at most $(\nu-1)j$.

Note that although before we have considered the degrees $\varkappa_j$ as integer numbers, the obtained result (\ref{IntKappa}) can be generalized to the case of non-integer $\varkappa_j$. We have
\begin{equation}\label{FunkEvolSimOP}
U_F(\tau; \bs{x}) = \frac{e^{-m^2 \tau} \tau^{-d/2\nu}}{(4\pi)^{d/2}} \sum\limits_{n} \frac{\gamma^n}{n!} \tau^{|n|-n\varkappa/\nu} \calE_{\nu,d/2}^{n\varkappa} \left( -\frac{x^2}{4\tau^{1/\nu}} \right).
\end{equation}
Relatively recently, analogous expressions for the case $d = 3$ applicable to models of the Ho\u{r}ava--Lifshitz type were obtained in the work \cite{Mamiya14}.

In particular, in the important ``coinciding points limit'', when $\bs{x} = 0$,
\begin{equation}
U_F(\tau; 0) = \frac{e^{-m^2 \tau} \tau^{-d/2\nu}}{(4\pi)^{d/2}\nu\Gamma(d/2)} \sum\limits_n \frac{\gamma^n}{n!} \Gamma\left(\frac{d/2+n\varkappa}{\nu}\right) \tau^{|n|-n\varkappa/\nu}.
\end{equation}

%%%%%%%%%%%%%%%%%%%%%%%%%%%%
\paragraph{Green functions of $O(d)$-invariant operators.}

Using the general method for calculating integrals of functions of the hypergeometric type \cite{Marichev}, the heat kernel (\ref{FunkEvolSimOP}) can be integrated over proper time, obtaining the general expression for the Green function of an arbitrary $O(d)$-invariant differential operator in the form of an infinite series in Fox $H$-functions. Indeed,
\begin{equation} \label{Grin1}
G_F(\bs{x}) = \int\limits_0^\infty U_F(\tau; \bs{x}) d\tau = \frac{1}{(4\pi)^{d/2}} \sum\limits_n \frac{\gamma^n}{n!} I_{|n|\nu-n\varkappa}^{n\varkappa}(x^2).
\end{equation}

In this case, the integrals $I_\alpha^\beta(x^2)$ can be represented as a Mellin convolution
\begin{equation}
I_\alpha^\beta(x^2) = \int\limits_0^\infty e^{-m^2\tau} \tau^\frac{-d/2+\alpha}{\nu} \calE_{\nu,d/2}^\beta\left(-\frac{x^2}{4\tau^{1/\nu}}\right) d\tau = \int\limits_0^\infty \phi_1(t)\phi_2(x^2/4t)\frac{dt}{t},
\end{equation}
where $\phi_1(t) = \nu e^{-m^2 t^\nu} t^{-d/2+\alpha+\nu}$ and $\phi_2(t) = \calE_{\nu,d/2}^\beta(-t)$. In the Mellin transform, the image of a convolution  is equal to the product of the images of the convolved functions ${I_\alpha^\beta}^*(s) = \phi_1^*(s)\phi_2^*(s)$, where
\begin{gather}
\phi_1^*(s) = m^{\frac{d-2\alpha-2s}{\nu}-2} \Gamma\left(\frac{-d/2+\alpha+s}{\nu}+1\right), \\
\phi_2^*(s) = \Gamma\left(\frac{d/2+\beta-s}{\nu}\right) \frac{\Gamma(s)}{\nu\Gamma(d/2-s)}.
\end{gather}
Then the value of the integral $I_\alpha^\beta(x^2)$ is given by the inverse Mellin transform
\begin{multline} \label{Grin2}
I_\alpha^\beta(x^2) = \frac{1}{2\pi i} \int\limits_C {I_\alpha^\beta}^*(s) (x^2/4)^{-s} ds = \\
m^{\frac{d-2\alpha}{\nu}-2} \frac{1}{2\pi i} \int\limits_C \frac{\Gamma(-s)\Gamma\left(\frac{-d/2+\alpha-s}{\nu}+1\right) \Gamma\left(\frac{d/2+\beta+s}{\nu}\right)}{\nu\Gamma(d/2+s)} \left(\frac{m^{2/\nu}x^2}{4}\right)^s ds = \\
m^{\frac{d-2\alpha}{\nu}-2} \frac{1}{\nu} H^{2,1}_{1,3}\left[\left.\frac{m^{2/\nu}x^2}{4}\right| \begin{smallmatrix}\left(1-\frac{d/2+\beta}{\nu},\frac{1}{\nu}\right) \\ (0,1), \left(1-\frac{d/2-\alpha}{\nu}, \frac{1}{\nu}\right), \left(\frac{d}{2},1\right)\end{smallmatrix}\right].
\end{multline}

%%%%%%%%%%%%%%%%%%%%%%%%%%%%%%%%%%
\section{Conclusion}
%%%%%%%%%%%%%%%%%%%%%%%%%%%%%%%%%%

We obtained an analytic expression for the heat kernel $U_{\nu, d}(\tau; \bs{x})$ of the operator $-(-\Delta)^\nu$, which is a generalization of (\ref{HeatKernel}), in terms of Fox--Wright psi functions. It should underline that its calculation was carried out by two essentially different methods. They are promising for further generalization to curved space (the first) and operators of a more complex type (the second).

For the first time we consider the general case (\ref{ConstCoeff}) of operators $F = -(-\Delta)^\nu + K(\nabla) - m^2$ and discuss the question of the behavior of their heat kernel $U_F(\tau; \bs{x})$ for small values of proper time $\tau$.

Asymptotic expressions for the functions $\calE_{\nu, \alpha}(z)$ for $z\to\infty$ also are obtained. They are necessary to study the asymptotic behavior of heat kernels and Green functions and the quantities constructed from them. The asymptotic expansion (\ref{Asympt}--\ref{Asympt2}) demonstrates the exponential behavior for integer $\nu$ and the power-law behavior for non-integer $\nu$, which is related to the local and, accordingly, non-local character of the operator $-(-\Delta)^\nu$ in these cases.

Using the Mellin transform (\ref{Mellin}) and the general method of integrating functions of the hypergeometric type presented in \cite{Marichev} we developed an algorithm for integrating expressions containing the functions $\calE^\beta_{\nu, \alpha}(z)$. The simple example (\ref {GrinNu}) demonstrates that the developed technique can be used as an alternative to the previously known methods of calculation. However, it is capable to significantly simplify the calculations and give results in more complex cases where the standard methods cannot be used.

Finally, we apply these general methods to the case (\ref{SphSymm}) of $O(d)$-invariant operators $F = -(-\Delta)^\nu + \sum \gamma_j (-\Delta)^{\varkappa_j} - m^2$ with fractial $\varkappa_i<\nu$ and obtain exact analytic expressions for the heat kernels (\ref{FunkEvolSimOP}) and Green functions (\ref{Grin1}, \ref{Grin2}) of these operators. As far as we know, expressions for the heat kernels for arbitrary dimension $d$ and general expressions for the Green functions of $O(d)$-invariant operators are given in the literature for the first time. The relative ease of obtaining them demonstrates clearly the capability of methods that have been used.

It is interesting  that the results obtained for $O(d)$-invariant operators remain valid not only for integer but also for non-integer values of the parameters $\nu$ and $d$. On the one hand, this makes it possible to use them in QFT both for dimensional regularization, when spaces of non-integer dimension are formally considered, and for allowing the introduction of differential operators of non-integer order. Such operators can be used to regularize Feynman diagrams in QFT, instead of introducing new terms with higher derivatives, by shifting the order of the differential operator in the kinetic term by a small amount such that $\nu = 2+\epsilon$. The possibility of such regularization in principal has recently been shown in \cite{Tarasov18} using the example of the massless $\phi^4$-theory. The prospects for the application of this new method will be discussed in our future papers.

On the other hand, the connection with fractional calculus opens the prospect of applying the obtained heat kernels far beyond the area of QFT. The theory of fractional differential equations can be effectively used to construct phenomenological models of fractal media, systems with memory and non-local interaction. As a result in recent years it has been increasingly used in a wide range of fields of physics, chemistry and biology --- in hydrodynamics and plasma physics, the theory of metals and semiconductors, polymers and nanomaterials, in the description of anomalous diffusion, high-temperature superconductivity, etc. We can talk about the rapid formation of a new interdisciplinary field and a special paradigm of research --- ``fractional dynamics''. Numerous applications of fractional calculus to physical problems are discussed, for example, in \cite{Tarasov} and the references there.

The wide application of these new methods to solve a variety of practical problems urgently requires the further development of computational methods. In connection with this it seems to us that the combination of the two previously disjointed areas --- fractional calculus and the heat kernel method --- can be extremely fruitful and requires careful study.

In the subsequent papers we use the asymptotic behavior of the functions $\calE_{\nu, \alpha}(z)$ and the technique of adiabatic expansion with respect to the dimension of background fields for generalization to the case of a Riemannian manifold. In particular, we define the analog of HaMiDeW-coefficients for higher-order operators and obtain recurrence relations on them. Such a generalization will allow us to develop an alternative method for calculating the effective action for theories with higher derivatives.

%%%%%%%%%%%%%%%%%%%%%%%%%%%%%%%%%%
\section*{Acknowledgements}
%%%%%%%%%%%%%%%%%%%%%%%%%%%%%%%%%%

The authors express their deep gratitude to the colleagues A.\,E.~Kazantsev, A.\,A.~Lobashev and M.\,M.~Popova for numerous and fruitful discussions of the material of the article, O.\,I.~Marichev for the instruction on the method of finding the asymptotic expansion of the Fox--Wright psi functions and  A.\,O.~Barvinsky for discussing the obtained results.

\bibliographystyle{unsrt}
\bibliography{HeatKernel}

\end{document}